\documentclass[twocolumn,aps,prb,showpacs,amsmath,superscriptaddress,longbibliography,notitlepage]{revtex4-1}
\usepackage{amssymb}
\usepackage{mathrsfs}
\usepackage{graphicx}
\usepackage{float}
\usepackage[caption=false]{subfig}
\usepackage[pdftex,colorlinks=red]{hyperref}
\usepackage{verbatim}
\usepackage{txfonts}
\usepackage{epstopdf}
\usepackage[normalem]{ulem}
\usepackage{xcolor}
\usepackage{bm}
\usepackage[utf8]{inputenc}

\usepackage{braket}
\usepackage{amsmath}

\usepackage{lipsum}

\definecolor{brightmaroon}{rgb}{0.92, 0.03, 0.08}

\def\LLH{\textcolor{cyan}}

\bibpunct{[}{]}{,}{n}{}{}

\begin{document}

\title{A brief review of hybrid skin-topological effect}
\author{Weiwei Zhu}
\affiliation{College of Physics and Optoelectronic Engineering, Ocean University of China, Qingdao 266100, China}
\affiliation{Qingdao Key Laboratory for Optics Photoelectronics $\&$ Engineering Research Center of Advanced Marine Physical Instruments and Equipment of Education Ministry, Ocean University of China, Qingdao 266100, China}
\author{Linhu Li}\email{lilh56@mail.sysu.edu.cn}
\affiliation{Guangdong Provincial Key Laboratory of Quantum Metrology and Sensing $\&$ School of Physics and Astronomy, Sun Yat-Sen University (Zhuhai Campus), Zhuhai 519082, China}

\date{\today}

\begin{abstract}
The finding of non-Hermitian skin effect has revolutionized our understanding of non-Hermitian topological phases, where the usual bulk-boundary correspondence is broken and new topological phases specific to non-Hermitian system are uncovered. Hybrid skin-topological effect (HSTE) is a class of newly discovered non-Hermitian topological states that simultaneously supports skin-localized topological edge states and extended bulk states. Here we provide a brief review of HSTE, starting from different mechanics that have been used to realize HSTE, including non-reciprocal couplings, onsite gain/loss, and non-Euclidean lattice geometries. We also review some theoretical developments closely related to the HSTE, including the concept of higher-order non-Hermitian skin effect, parity-time symmetry engineering, and non-Hermitian chiral skin effect. Finally, we summarize recent experimental exploration of HSTE, including its realization in electric circuits systems, non-Hermitian photonic crystals, and active matter systems. We hope this review can make the concept of hybrid-skin effect clearer and inspire new finding of non-Hermitian topological states in higher dimensional systems.
\end{abstract}

\maketitle
\tableofcontents

\section{introduction}
During the past two decades, non-Hermitian physics has been one of the most active and rapidly advancing branches of physics, where effective non-Hermitian Hamiltonians~\cite{nonH1,nonH2} are introduced to describe non-conservative systems with particle or energy gain and loss~\cite{rotter2009non,yoshida2018non,shen2018quantum,yamamoto2019theory,ma2016acoustic,cummer2016controlling,zangeneh2019active,feng2017non,el2018non,longhi2018parity,ozawa2019topological}.
In the early years, investigations of non-Hermitian physics mainly focus on aspects, namely parity–time ($PT$) symmetry~\cite{nonH1,feng2017non,el2018non}, and exceptional points (EPs)~\cite{rotter2009non,kato2013perturbation,heiss2004exceptional}.
Representing a specific type of the more general concept of pseudo-Hermiticity~\cite{mostafazadeh2002pseudo},
$PT$ symmetry ensures a system to have real eigenenergies in the so-called $PT$-unbroken phase, or paired ones with opposite imaginary parts in the $PT$-broken phase.
On the other hand, EPs are a special type of degeneracy in non-Hermitian systems, where two or more eigenstates become the same not only in their eigenenergies, but also the states themselves.
This is commonly known as the coalescence of eigenstates of rank-deficient Hamiltonian matrices.
The spectral singularity of EPs not only signals the transition between $PT$-unbroken and broken phases, but also leads to other intriguing properties of non-Hermitian systems,
such as unidirectional invisibility~\cite{lin2011unidirectional,feng2013experimental}, enhanced sensitivity~\cite{wiersig2014enhancing,liu2015metrology,hodaei2017enhanced,chen2017exceptional},
and unusual topological properties along trajectories encircling EPs~\cite{dembowski2001experimental,gao2015observation,mailybaev2005geometric,lee2016anomalous,leykam2017edge,shen2018topological,yin2018geometrical,li2019geometric,hu2017exceptional,hassan2017dynamically}


More recently, many efforts have been devoted to investigate how non-Hermiticity affects topological phases of matter~\cite{hasan2010colloquium,qi2011topological,yan2017topological,wen2017colloquium}, another subfield of physics that has received enormous attention over the past half-century.
In studying topological physics of Hermitian systems, bulk-boundary correspondence (BBC) is a fundamental principle that determines the number of topological edge states by a class of global quantities called topological invariants.
In some non-Hermitian systems, conventional BBC is broken by the massive accumulation of eigenstates to the boundaries~\cite{yao2018edge,martinez2018non}, known as the non-Hermitian skin effect (NHSE)~\cite{yao2018edge}. General Brillouin zone and non-Bloch band theory~\cite{yao2018edge,yokomizo2019nonB,yang2020nonH} are developed to correctly characterize the non-Hermitian topological edge states by non-Bloch topological invariants.
In addition, NHSE itself also represents a boundary localization corresponding to a spectral winding topology originating from the complex eigenenergies of non-Hermitian Hamiltonians~\cite{borgnia2020non,okuma2020topological,zhang2020correspondence,li2021quantized}.
Since its discovery, NHSE has triggered an explosively growing number of extensive investigations into its various aspects,
with the latest developments summarized in several recent reviews~\cite{ding2022non,zhang2022review,lin2023topological,banerjee2023non,okuma2023non,Zhou2023floquet}.

Manifesting as a boundary localization of eigenstates, NHSE becomes more sophisticated in higher dimensional systems, where richer structures of boundaries emerge from different geometries~\cite{zhang2022universal,Wang2022_NC,Fang2022,zhou2023observation,wan2023,wang2023experimental,qin2023geometry} and defects~\cite{disclination_NHSE,dislocation_NHSE1,discloation_NHSE2,dislocation_NHSE3,manna2023inner} of lattices.
In this topical review, we provide an overview of the discovery and developments of HSTE~\cite{Lee2019hybrid}, a novel phenomenon arising from the interplay of NHSE and topological boundary states in two- or higher-dimensions.
A characteristic feature of HSTE is that the NHSE acts only on boundary states, inducing localized eigenstates at lower dimensional boundaries, while leaving bulk states extended in the system.
The emergence of HSTE was first proposed in a non-reciprocal square lattice~\cite{Lee2019hybrid},
and later in other lattices with on-site gain and loss~\cite{li2020topological,li2022gain,zhu2022hybrid} or non-Euclidean geometry~\cite{sun2023hybrid}, as we will discuss in detail in Sec. \ref{sec:mechanics}.
In Sec. \ref{sec:beyond}, we introduce
several related theoretical concepts and phenomena that provide explanation and engineering methods of HSTE based on different mechanisms, including
higher-order NHSE~\cite{PhysRevB.102.205118}, selective activation of bulk and boundary NHSE~\cite{lei2023pt}, and boundary NHSE with Hermitian bulks~\cite{ma2023nonH,schindler2023hermitian,daichi2023universal}.
Beyond theoretical exploration, HSTE has also been realized in several experimental setups of circuit lattices, photonic crystals, and active particles, which will be reviewed in Sec. \ref{sec:exp}.

Before moving on to the next section, we first provide a brief introduction of NHSE to set the stage for discussing its hybridization with topological boundary localization.
NHSE represents an intriguing phenomenon of non-Hermitian Hamiltonians, where most (if not all) eigenstates localize at the boundaries of the systems.
The terminology of ``NHSE" was first proposed by Yao and Wang in 2018, in a seminal theoretical study that settles the problem of topological  BBC in one-dimensional non-Hermotiian systems~\cite{yao2018edge}.
On the other hand, such a massive boundary accumulation of eigenstates for non-Hermitian Hamiltonians has also been noticed in another contemporaneous work~\cite{martinez2018non}, and intensively investigated by mathematicians decades ago as a property of banded Toeplitz matrices~\cite{bottcher2005spectral}.
\begin{figure}
\includegraphics[width=1\linewidth]{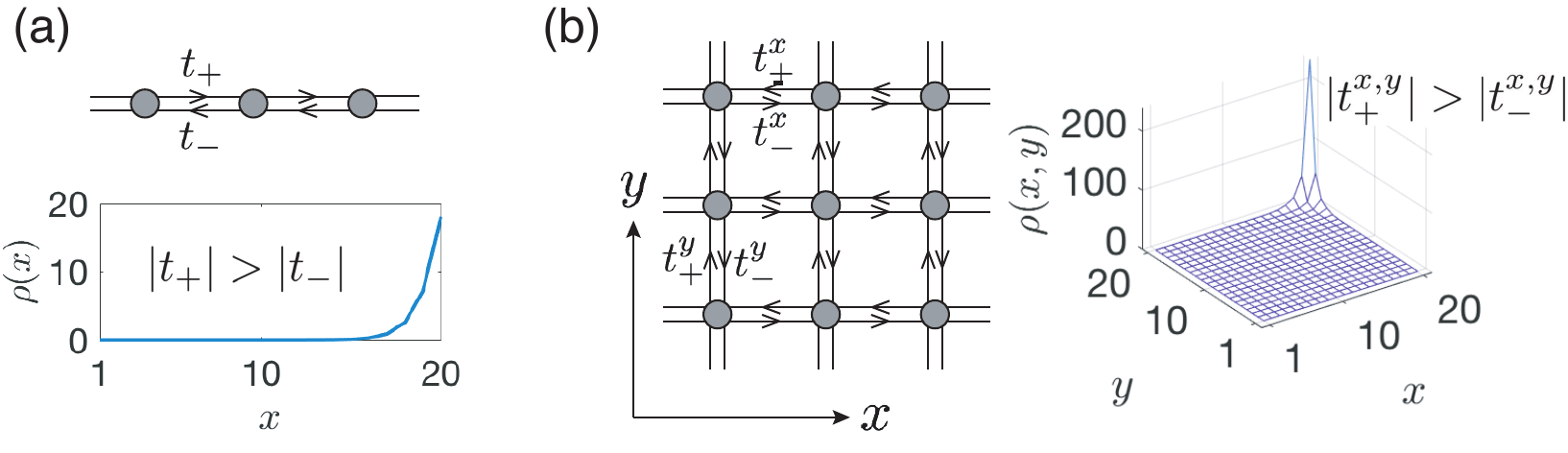}
\caption{(a) The Hatano Nelson model with non-reciprocal hopping $t_+\neq t_-$, and the summed squared amplitudes over all eigenstates, which accumulate to the edge of the 1D system.
(b) A 2D Hatano-Nelson model with non-reciprocal hopping $t^{x,y}_+\neq t_-^{x,y}$ and the summed squared amplitudes over all eigenstates in the 2D lattice, which show a clear corner skin localization.
Reproduced from Ref. \cite{Lee2019hybrid}.}
\label{fig:NHSE}
\end{figure}

The simplest example for demonstrating NHSE is the single-band Hatano-Nelson model with asymmetric hopping amplitude~\cite{HN1,HN2}, described by the Hamiltonian
\begin{eqnarray}
H=\sum_{x}t_+\hat{c}^\dagger_{x+1} \hat{c}_{x}+t_-\hat{c}^\dagger_{x} \hat{c}_{x+1},
\end{eqnarray}
with $\hat{c}_x$ the annihilation operator of a particle at site $x$. Under non-reciprocal real positive hopping parameters $t_+ \neq t_-$, all  eigenstates are found to be localized toward the direction of the stronger hopping, as shown in Fig. \ref{fig:NHSE}(a).
Note that these eigenstates are usually still referred as ``bulk" states as their number scales with the system's size.
In this simple model, the NHSE under OBC can be unveiled through a similarity transformation of the Hamiltonian matrix. That is, we can obtain a Hermitian matrix $\bar{H}$ through $$\bar{H}=S^{-1}HS,$$ with $S={\rm diag}[1,r,r^2,...r^{L-1}]$ and $r=\sqrt{t_+/t_-}$.
In this way, eigenstates $|\bar{\psi}\rangle$ of $\bar{H}$ must be extended,
thus the original Hamiltonian has exponentially localized eigenstates given by $|\psi\rangle=S|\bar{\psi}\rangle$.
A straightforward extension into higher dimensions is to consider a 2D version of the Hatano-Nelson model, with non-reciprocal hoppings along both directions. Consequently, a corner NHSE arises with all eigenstates are now localized at a corner of the 2D lattice, as shown in Fig. \ref{fig:NHSE}(b).

Note that for more complicated non-Hermitian systems, e.g. with multiple bands or long-range hoppings, it may be impossible to find such a similarity transformation to map the system to a Hamiltonian without NHSE. This is because generally speaking, eigenstates can exhibit NHSE with different localizing strength and directions.
A comprehensive description of NHSE requires a complex analytical continuation of the 1D Bloch momentum, known as the non-Bloch band theory~\cite{yao2018edge,yokomizo2019nonB,yang2020nonH}, or intruducing a geometrical object known as the amoeba for two- or higher-dimensional systems~\cite{wang2022amoeba,hu2023non},
which are beyond the scope of this review.

\section{HSTE from different mechanics}\label{sec:mechanics}
\subsection{HSTE induced by non-reciprocal hopping}
\begin{figure*}[htbp]
\includegraphics[width=1\linewidth]{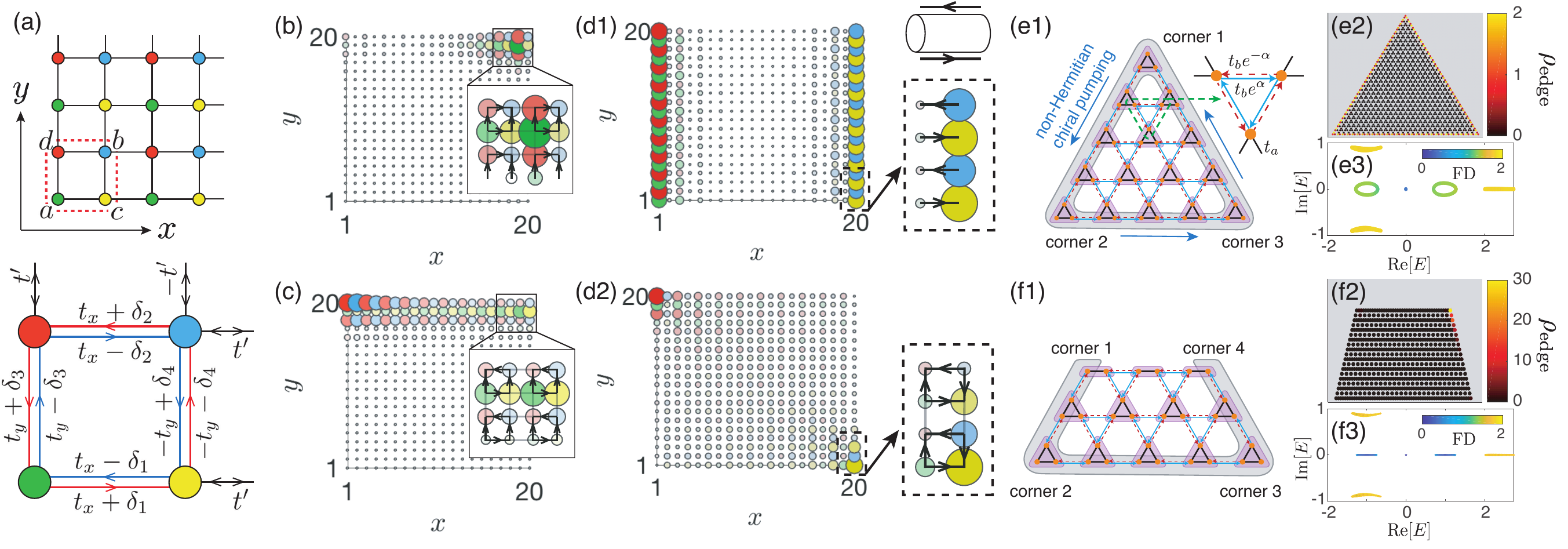}
\caption{
HSTE and boundary spectral winding in 2D non-reciprocal lattices.
(a) A square lattice with four sublattices in each unit cell, and non-reciprocal real hopping parameters $\delta_{1,2,3,4}$ along the two directions.
(b) Corner NHSE in the OBC square lattice, with $\delta_1\neq\delta_2$ and $\delta_3\neq\delta_4$.
(b) line NHSE in the OBC square lattice, with $\delta_1=\delta_2$ and $\delta_3\neq\delta_4$.
(d1) Emergence of 1D boundary modes under x-OBC/y-PBC (cylinder geometry), with $\delta_1=\delta_2$ and $\delta_3=\delta_4$.
(d2) hybrid skin-topological corner states emerge in the OBC square lattice with destructive interference of bulk non-reciprocity.
(e) A non-Hermitian breathing Kagome lattice in a triangular geometry, with three sublattices in a unit cell, where non-reciprocity described by $\alpha$ is balanced between the three intercell hoppings (blue solid and red dash lines). Gray area in (e1) indicates the effective 1D boundary system.
(e2) and (e3) display the summed distribution of edge states and the complex spectrum of the system.
(f) the same as (e), but with a trapezoidal geometry.
Reproduced from Refs. \cite{Lee2019hybrid,ou2023nonH}.
}
\label{non_reciprocity}
\end{figure*}
Early studies on NHSE and topological phases mostly focus on how the former alternates the BBC of the later.
In contrast, HSTE represents a different interplay between NHSE and conventional topological phases, where they are treated on equal footing to give rise to eigenstate localization at lower-dimensional boundaries (e.g., corners in 2D lattices).
The concept of HSTE was first proposed by Lee, Li, and Gong in 2019~\cite{Lee2019hybrid}, using a 2D extension of the 1D non-Hermitian Su-Schrieffer-Heeger (SSH) model~\cite{SSH} with NHSE. Specifically, the model is constituted by two copies of the 1D model along each direction, described by four different non-reciprocal parameters ($\delta_{1,2,3,4}$), as shown in Fig. \ref{non_reciprocity}(a).
NSHE naturally arises in this model under the OBCs, yet shows distinguished behaviors in different parameter regimes.
Namely, a corner NHSE general emerges for all bulk eigenstates when there is no specific restriction of parameters~[Fig. \ref{non_reciprocity}(b)],
and a line NHSE emerges when the non-reciprocity is balanced and destructively interferes along one direction~[Fig. \ref{non_reciprocity}(c)].
Finally, in the parameter regime where non-reciprocities cancel in both direction,
no skin effect is observed for bulk states in either direction, yet corner NHSE is still seen under full OBCs.
This is because this model also support first-order topological states, which possess unequal distribution on different sublattices~[Fig. \ref{non_reciprocity}(d1)].
Consequently, these states do not experience the full destructive interference of non-reciprocity,
exhibiting NHSE and thus giving rise to the new class of hybrid skin-topology corner states~[Fig. \ref{non_reciprocity}(d2)].
A distinctive feature of these corner states is that their number scales as $\mathcal{O}(L)$ with $L$ the system length,
in contrast to the corner NHSE scaling as $\mathcal{O}(L^2)$ or higher-order topological corner states scaling as $\mathcal{O}(1)$.
In higher dimensions, richer classes of HSTE can be generated by considered different skin or topological localization along each direction,
where the number of hybrid skin-topology states scales as
$\mathcal{O}(L^S)$, with $S$ the number of directions affected by the NHSE.

A key factor for the emergence of HSTE in the above mentioned model is that two neighbor edges possess non-reciprocity toward the same corner. In a recent study, it has been shown that a non-Hermitian breathing Kagome lattice also
possesses both the destructive interference of non-reciprocity and first-order edge states,
yet it support a different type of boundary phenomenon dubbed as the boundary spectral winding~\cite{ou2023nonH}.
As shown in Fig.~\ref{non_reciprocity}(e1), a triangle geometry of this lattice has its three boundary constituting a closed 1D system with chiral non-reciprocity.
Thus first-order edge states remain extended along the boundaries, yet their eigenenergies form a loop-like spectrum in the complex energy plane [Fig.~\ref{non_reciprocity}(e2) and (e3)].
This observation is analogous to a 1D non-reciprocal lattice, where NHSE under OBCs can be related to a nontrivial winding of loop-like complex spectrum under PBCs~\cite{spectralwinding1,spectralwinding2,spectralwinding3}.
In a trapezoidal lattice, on the other hand, the top boundary possesses the opposite chirality to the rest three boundaries,
which constitute an effectively open 1D non-reciprocal system~[Fig.~\ref{non_reciprocity}(f1)]. Thus HSTE emerges provided the top boundary has a length comparable with the rest, and the loop-like boundary spectrum collapses in to some lines~[Fig.~\ref{non_reciprocity}(f2) and (f3).

\subsection{HSTE induced by gain and loss}
\begin{figure*}[htbp]
\includegraphics[width=0.75\linewidth]{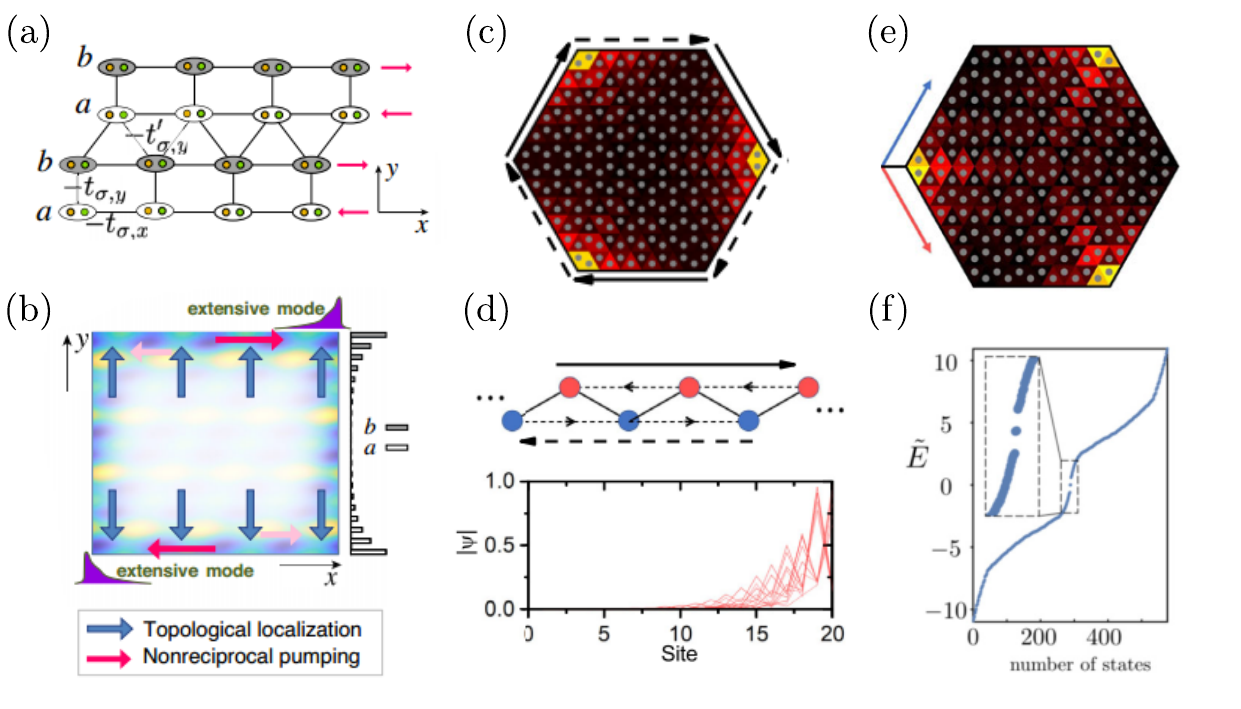}
\caption{
Hybrid skin-topological effect by gain and loss.
(a) A coupled-wire construction of HSTE by loss. Each unit cell contains sublattices (a and b) and pseudospins (yellow and green sites). The pink arrows mark the direction of nonreciprocity of each sublattice. (b) Mechanics for the formation of corner skin modes. Along $y$ direction, the boundary states are localized due to topological localization and localized at a (b) sublattice at lower (upper) boundary. The nonreciprocity along $x$ direction makes the boundary states pump to corners. (c) Corner skin modes in non-Hermitian Haldane model as accumulation of chiral edge states. Solid (dashed) lines mark the gain (loss) edge states. (d) An intuitive understanding of corner skin modes as the NHSE of zigzag chain at boundary. (e) The field distribution of corner skin modes in $\mathcal{PT}$ symmetry broken phase. Red (green) arrows mark the gain (loss) edge. The field prefers to distribute at gain edge. (f) Spectrum of a higher-order topological insulator constructed by extending the non-Hermitian Haldane model. The existence of topological corner modes in the extended Hermitian Hamiltonian is an evidence of the existence of HSTE in non-Hermitian Haldane model. Reproduced from Refs. \cite{li2020topological,li2022gain,zhu2022hybrid}.}
\label{loss}
\end{figure*}

Since its discovery, many works have proposed to construct HSTE solely by non-Hermitian gain and loss, which are considered to be experimentally easier to implement than non-reciprocal couplings. The first work is given by Linhu Li, Ching Hua Lee and Jiangbin Gong, in a cold-atom system based on a coupled-wire construction~\cite{li2020topological}. They consider an anisotropy two-dimensional tight-binding model, which is composed of a class of $x$-direction chains and are coupled by dimerized couplings along $y$ direction [Fig.~\ref{loss}(a)]. The $x$-direction chains support NHSE due to the specific design of couplings and introduction of atomic loss. The $y$-direction couplings make the system behave as SSH model along that direction. Each unit cell contains both a sublattice degree of freedom and a pseudospin degree of freedom. The loss is introduced to both sublattices but only in one pseudospin. The resultant NHSE in different sublattices makes eigenstates favoring to localize toward opposite directions. Therefore they cancel each other in the bulk by the inter-sublattice couplings, so that the bulk states remain extended [Fig.~\ref{loss}(b)]. Along the edges, however, one sublattice is isolated due to the topological nontrivial phase of SSH model, so that the NHSE is kept and makes the edge states localized at corners, which are called corner skin modes. On the other hand, the HSTE can also be described by a non-zero Chern number of the bulks bands, and can be switched on and off by changing the bulk topology.

Later in 2022, the HSTE has been studied in the Haldane model~\cite{Haldane1988} with non-Hermitian gain and loss by two different groups~\cite{li2022gain,zhu2022hybrid}. In this model, the corner skin modes can be understood as the accumulation of chiral edge states with net gain and loss [Fig.~\ref{loss}(c)]. These corner modes appear only at the connections of zigzag edges, whose net gain and loss are non-zero. Along the armchair edge, the gain and loss are balanced with each other so the chiral edge states are still extended. Ref.~\cite{li2022gain} points out that the corner skin modes can also be understood as coupled zigzag chains with opposite NHSE in odd and even layers [Fig.~\ref{loss}(d)]. In the bulk, the opposite directions of NHSE cancel each other so the bulk states are extended. At edge, the net gain/loss and the nonlocal edge currents make the chiral edge states localized at corners. In addition, they also found a $\mathcal{PT}$ phase transition of the skin topological modes. In the $\mathcal{PT}$ symmetric phase, the corner skin modes symmetrically distributed along the gain edge and loss edge. When the gain/loss are large enough, some corner skin modes are located at $\mathcal{PT}$ symmetry broken phase and they distribute more on the gain edge [Fig.~\ref{loss}(e)].

In comparison, Ref.~\cite{zhu2022hybrid} has focused on several different aspects to understand HSTE in the non-Hermitian Haldane model.
As a special kind of second-order NHSE (see Sec. \ref{sec:hoNHSE} for further discussion),
the HSTE here exhibit several interesting properties compared with other types of NHSE.
i) For a $L\times L$ system, there are $\mathcal{O}(L)$ eigenmodes localized at corners, while the bulk states are still extended. Such localization is different from the higher-order topological corner modes, where $\mathcal{O}(1)$ modes localize at corners, and also different from the corner skin modes from first-order NHSE, where $\mathcal{O}(L\times L)$ modes localize at corners.
ii) The spectrum of topological edge states are sensitive to boundary conditions. Under OBCs in one direction and PBCs in another direction, the spectrum of topological edge states is complex and encircles an area. Under OBCs in both directions, edge spectral collapses to real values. Meanwhile, bulk spectrum remains unchanged for different boundary conditions.
Note that these two features of HSTE have also been revealed in the first proposal of HSTE \cite{Lee2019hybrid}.
iii) By extending the Hamiltonian of non-Hermitian Haldane model to an extended Hermitian Hamiltonian with chiral symmetry, the HSTE is connected to a second-order topological insulator [Fig.~\ref{loss}(f)].
The corresponding higher-order topology can either be intrinsic (protected by bulk gap) or extrinsic (protected by edge gap), depending on extra symmetries of the extended Hamiltonian.
In particular,
The intrinsic-type HSTE is pseudo-inversion symmetric, which ensures the corresponding extended Hermitian Hamiltonian being inversion-symmetric. With further symmetry analysis of the extended Hamiltonian, they found that it has no Wannier representation but can be Wannierized by adding a trivial insulator, which means it is a fragile topological insulator supporting topological corner states. In the extrinsic-type HSTE, the pseudo-inversion symmetry is broken by  a on-site potential difference (for two sublattices),
and the corresponding extended Hamiltonian support a transition between topologically trivial and nontrivial phases without closing the bulk band gap. Therefore HSTE can be switched off by the on-site potential difference.
In addition, the HSTE is quite general and can also be extended to nonequilibrium systems, such as periodically driven systems~\cite{zhu2022hybrid,Li2023floquet,Sun2023floquet}.
\begin{figure*}
\includegraphics[width=1\linewidth]{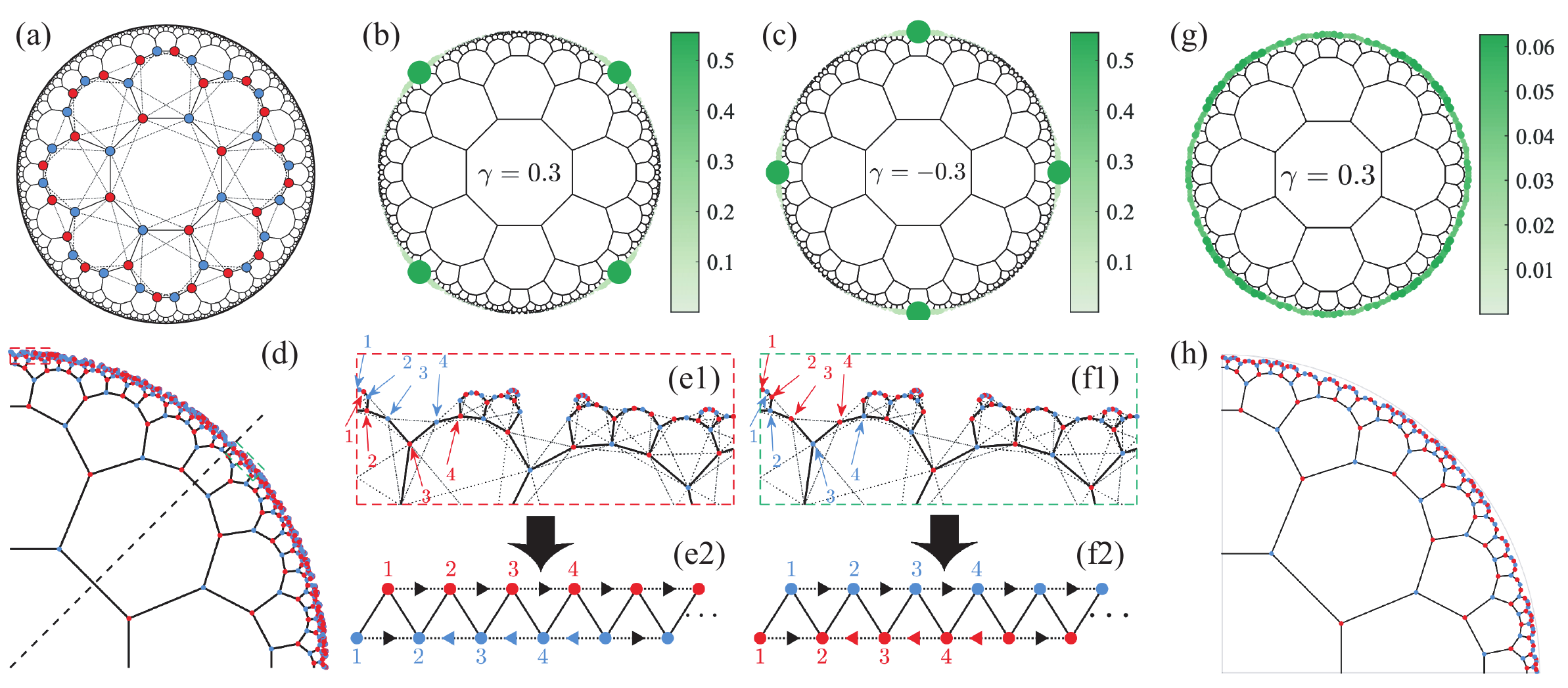}
\caption{(a) The hyperbolic Haldane lattice constructed from the contact of three octagons. Red and blue dots denote the two sublattices with on-site gain and loss $\pm i\gamma$; solid and dash lines denote the nearest and next-nearest hoppings, respectively, where the later possess sublattice-dependent phases $\phi$ as in the Haldane model.
(b) and (c) Distribution of skin-topological corner modes with different on-site gain and loss.
(d) Illustration of a quarter of the OBC hyperbolic lattice with complete unit cells. (e) and (f) Zoomed details of the boundary near the top and the diagonal, respectively, with their effective 1D structures displayed in the bottom. Arrows indicate the directions with positive phase factor $\phi$.
(g) Distribution of boundary states in the OBC hyperbolic lattice with a smoother boundary created by cutting the lattice with a fixed-radius circle.
(f) Illustration of a quarter of lattice in (g).
Reproduced from Ref. \cite{sun2023hybrid}.
}
\label{hyperbolic}
\end{figure*}

The above mentioned works have studied HSTE in lattice models.
Recently, the concept of HSTE has been extended to continuous systems by introducing pseudo-inversion symmetric gain and loss to photonic crystals with nontrivial Chern topology~\cite{Zhu2023}.
The HSTE in non-Hermitian photonic crystal was analyzed by extracting the topological edge states. Without gain and loss, the Hamiltonian of topological edge states can be described by a diagonal Hamiltonian under the basis of $(\psi_{\mathrm{up}},\psi_{\mathrm{down}})^\mathrm{T}$. The Hamiltonian of these topological edge states can be described by
\begin{equation}
  H_{\mathrm{edge}}=\omega_{r}\sigma_{0}+v(k_x-k_r)\sigma_z.
\end{equation}
Here $\omega_r$ is a reference energy frequency, $v$ is the group velocity, $k_x$ is the quasi-momentum along $x$ direction, and $k_r$ is a reference quasi-momentum. $\sigma_0$ and $\sigma_z$ represent the two-by-two identity matrix and the third Pauli matrix, respectively. WIth gain and loss introduced in a pseudo-inversion symmetric manner, the up edge states and down edge states have opposite gain and loss, and the Hamiltonian becomes
\begin{equation}
  H_{\mathrm{edge}}^{m}=\omega_{r}\sigma_{0}+v(k_x-k_r)\sigma_z+i\gamma\sigma_z.
\end{equation}
Therefore, the symmetric gain and loss can be treated as modification of the quasi-momentum to a complex value, $\tilde{k}_x=k_x+i\gamma/v$, which makes the topological edge states localized at corner under OBCs.
That is, the HSTE in non-Hermitian photonic crystals can also be understood as NHSE of one dimensional topological edge states.
It shares the same properties as HSTE for lattice models, e.g.,
the number of corner skin modes is proportional to $\mathcal{O}(L)$, and the spectrum of topological edge states is sensitive to the boundary conditions. Different localization phenomena, like frequency-position locked corner skin modes and multiple-corner skin modes, have also been uncovered in the non-Hermitian photonic crystals.

\subsection{HSTE in hyperbolic lattices}
Hyperbolic lattices are an exotic class of lattices in non-Euclidean space with constant negative curvature, supporting tessellation of any regular $p$-site polygon with $p>2$,
which has been recently realized in circuit quantum electrodynamics and electric circuits~\cite{kollar2019hyperbolic,lenggenhager2022simulating,zhang2022observation,zhang2023hyperbolic,chen2023hyperbolic}.
In Ref. \cite{sun2023hybrid}, it has been shown that HSTE can also emerge in hyperbolic Haldane models with sublattice-dependent gain and loss. An example with $p=8$ is sketched in Fig. \ref{hyperbolic}(a), where red and blue dots denote different sublattices with different on-site gain and loss $\pm i\gamma$.
Under the disk geometry, skin-topological corner states are found to appear at four diagonal corners and four axial corners, for positive and negative $\gamma$ respectively, as displayed in Fig. \ref{hyperbolic}(b) and (c).
The mechanism behind it is similar to the non-Hermitian Haldane model in Euclidean space~\cite{li2022gain,zhu2022hybrid}.
Namely, as seen in Fig. \ref{hyperbolic}(d) to (f), the hyperbolic lattice with complete unit cells possess a zigzag-like 1D boundary with non-uniform phase factors for hoppings, and hence nonzero net fluxes, near the above mentioned eight corners.
Its interplay with the sublattice-dependent gain and loss
leads to different skin accumulating directions for different segments of the boundaries,
and skin-topological corner modes emerge at the connections between them.
It is also noted that since the boundary geometry plays a crucial role for the appearance of skin-topological corner modes,
they may disappear in the same hyperbolic lattice but with a smoother boundary that does not support any net flux, as shown in Fig. \ref{hyperbolic}(g) and (h).
Similar results have also been discussed in this work for the case with $p=12$, where skin-topological states appear at six corners of the disk geometry due to the sixfold rotation symmetry.



\section{Related theoretical developments}\label{sec:beyond}

\subsection{higher-order NHSE}\label{sec:hoNHSE}

\begin{figure}[htbp]
\includegraphics[width=0.9\linewidth]{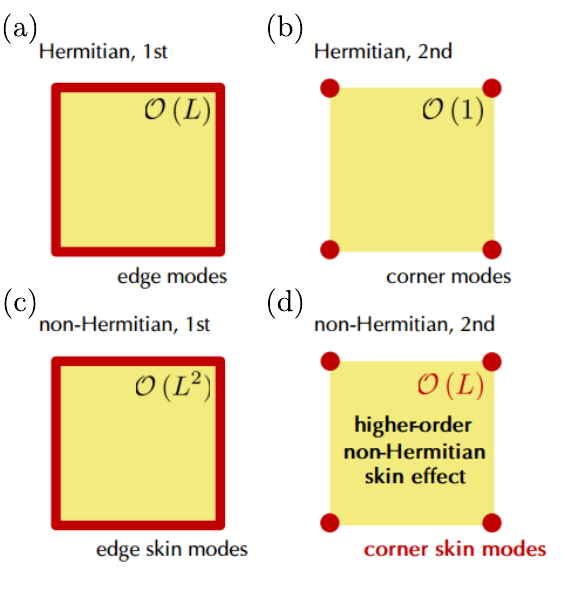}
\caption{
Diagram of different topological states in two dimensional systems.
(a) First-order topological insulator. (b) Second-order topological insulator. (c) First-order NHSE. (d) Second-order NHSE. 
Reproduced from Ref. \cite{PhysRevB.102.205118}.
}
\label{HONHSE}
\end{figure}

The HSTE, whose corner skin modes comes from the interplay of topological localization and NHSE,
is considered as a special kind of higher-order NHSE. The definition of higher-order NHSE was given by Kohei Kawabata, Masatoshi Sato and Ken Shiozaki~\cite{PhysRevB.102.205118}. According to the definition, $m^{th}$ order NHSE in $n$-dimensional system with system size $L$ has $\mathcal{O}(L^{n-m+1})$ states localized at corners. They use two dimensional system as an example to compare different Hermitian and non-Hermitian topological states with higher-order NHSE. For the first-order topological insulator, there are $\mathcal{O}(L)$ modes localized at edge [Fig.~\ref{HONHSE}(a)]; for the second-order topological insulator, there are $\mathcal{O}(1)$ modes localized at corners [Fig.~\ref{HONHSE}(b)]; for the first-order NHSE, there are $\mathcal{O}(L^2)$ modes localized at edge [Fig.~\ref{HONHSE}(c)]; however, for second-order NHSE, there are $\mathcal{O}(L)$ modes localized at corners [Fig.~\ref{HONHSE}(d)], which is different from all previous case. Almost the same time, another work also use the terminology of second-order NHSE~\cite{PhysRevB.102.241202}. They have connected the second-order NHSE with intrinsic second-order topological insulator and extrinsic second-order topological insulator.

Despite being widely used in literature, we note that both the terminologies of HSTE and higher-order NHSE have their own limitations for describing NHSE in higher-dimensional systems.
For “HSTE", as pointed out in Ref.~\cite{Lee2019hybrid}, skin eigenstates scales as $\mathcal{O}(L^{n-m+1})$ with $m>1$ do not necessarily require protection from band topological properties.
Nonetheless, these states generally originate from NHSE acting on conventional boundary states in periodic lattices, which in principle can be related to certain topology in a projected space (e.g., see Ref.~\cite{mong2013edge} for edge states in Dirac Hamiltonians).
On the other hand, ``higher-order NHSE" emphasizes the different scaling of the number of skin modes to the system's size,
but it does not distinguish bulk NHSE toward boundaries of different dimensions, e.g., corner and line NHSE in 2D.
For example,
corner and line NHSE are both first-order NHSE according to the definition here.
In the context of HSTE that treats NHSE and topological localization on equal footing,
they are classified as ``SS" and ``S0" phases respectively, with HSTE being the ``ST" phase.
 In addition, corner NHSE has also been dubbed as higher-order NHSE in some early studies~\cite{Lee2019hybrid,PhysRevB.103.045420}, since it represents a localization on higher-order boundaries solely induced by NHSE.

\subsection{Intrinsic HSTE through $PT$ engineering}
In 1D systems, a correspondence has been established between the emergence of NHSE under OBCs and a nontrivial spectral winding for eigenenergies under PBCs~\cite{spectralwinding1,spectralwinding2,spectralwinding3}.
In Ref.~\cite{lei2023pt}, this correspondence has been applied to design higher-dimensional models that possess HSTE, as well as some other types of NHSE selectively activated for bulk or boundary states.
The model engineering involves two steps:
\begin{itemize}
\item[(i)] designing a lower-dimensional non-Hermitian Hamiltonian with parity-time ($PT$) symmetry acting differently on bulk and boundaries; and
\item[(ii)]  introducing an extra crystal momentum that leads to nontrivial spectral winding for the $PT$-asymmetric states.
\end{itemize}
Thus, NHSE arises for these states when OBCs are taken along the extra direction.

\begin{figure}[htbp]
\includegraphics[width=1\linewidth]{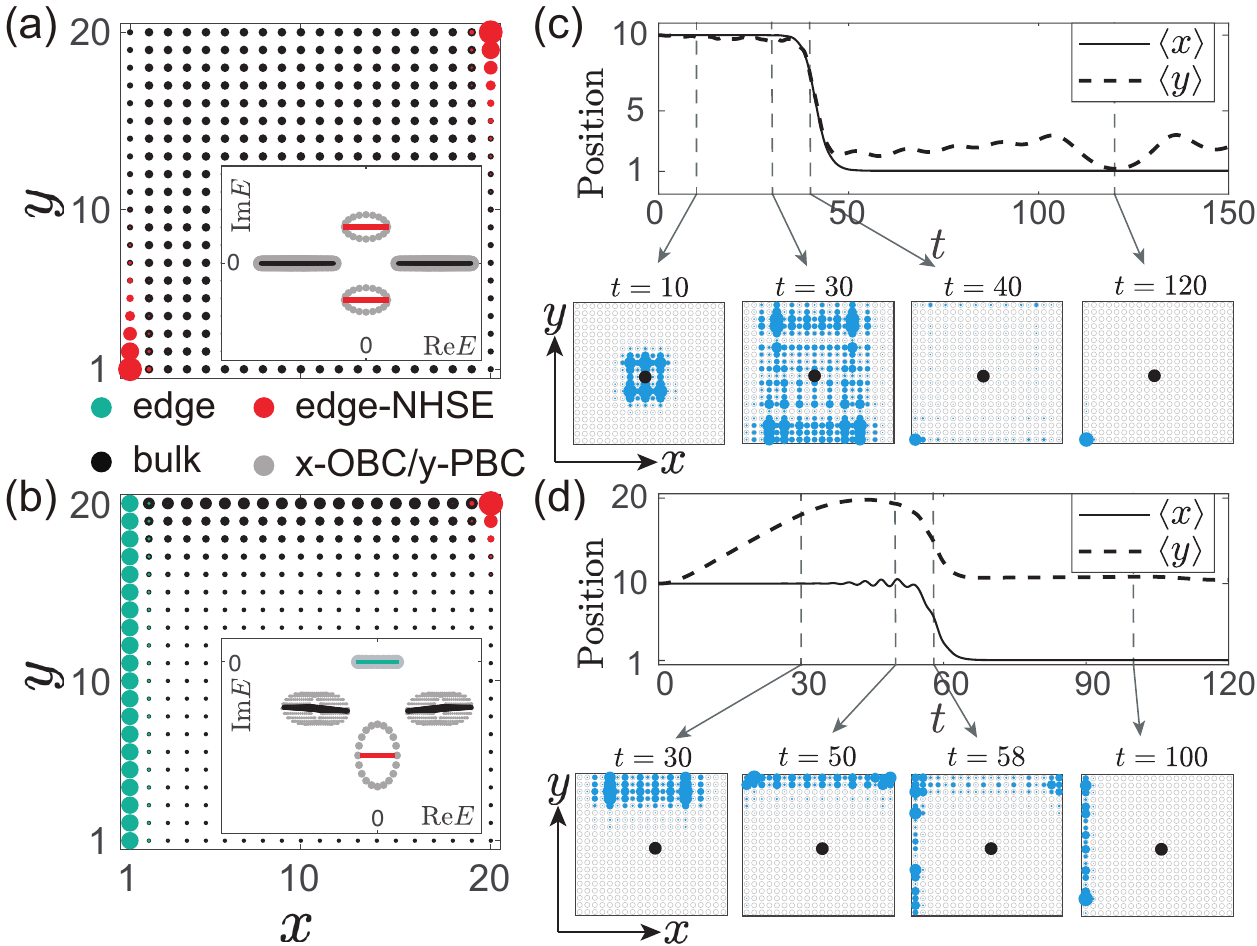}
\caption{
NHSE selectively activated for bulk or boundary states.
(a) Eigenstate distribution and eigenenergy spectrum for the intrinsic HSTE, where the absence of bulk NHSE is geometry-independent.
(b) Eigenstate distribution and eigenenergy spectrum when NHSE is turned on and off for bulk and one branch of edge states, respectively.
(c) and (d) The dynamical evolution of a center-localized initial state (black dots) for the model of (a) and (b), respectively.
Reproduced from Ref.~\cite{lei2023pt}.
Non-monotonic dynamics is observed in (d) from $\langle y \rangle$, the center-of-mass along $y$ direction.
}\label{PT_selective}
\end{figure}

In particular, an ``intrinsic" type of HSTE can be engineered based on the SSH model with sublattice-dependent gain and loss, described by the Hamiltonian
$$H_g(k_x)=(u+v\cos k_x)\sigma_x+v\sin k_x \sigma_y+ig\sigma_z,$$
with $\sigma_{x,y,z}$ the Pauli's matrices.
This Hamiltonian satisfies the $PT$ symmetry $\sigma_x H_g \sigma_x=H^*_g$, but possess $PT$-broken topological edge states (due to their sublattice polarization) under the OBCs.
Next, a 2D Hamiltonian is constructed as
$$H_{g,{\rm 2D}}=H_g(k_x)+t_1\cos k_y\sigma_0+t_2'\sin k_y\sigma_z$$
with $\sigma_0$ the two-by-two identity matrix.
In the parameter regime chosen for Fig. \ref{PT_selective}(a), bulk states are in the $PT$-unbroken phase and have purely real eigenenergies.
The resultant line-spectrum along the real axis cannot contain any nonzero spectral winding, suggesting the absence of bulk NHSE for arbitrary OBC geometries.
This is considered as an ``intrinsic" property of the model, as higher-dimensional non-Hermitian Hamiltonians generally have their bulk spectra covering a nonzero area on the complex energy plane, meaning that bulk NHSE shall emerge for generic OBC geometries~\cite{zhang2022universal}.
On the other hand, edge states in this model are $PT$-broken, and possess a loop-spectrum when the extra $k_y$ terms are introduced.
Therefore they exhibit $y$-NHSE when OBCs are taken accordingly, resulting in the so-called intrinsic HSTE.

In general, this $PT$-engineering method is useful for activating/deactivating NHSE on different bulk and boundary sectors for models of different dimensions.
For instance, in the above 2D model, NHSE can be turned on and off for bulk and one branch of edge states, respectively, by introducing an extra anti-Hermitian term $-i(t_2'\sin k_y+g)\sigma_0$, as shown in Fig. \ref{PT_selective}(b).
These phenomena manifest as distinguished dynamics in different stages during the evolution [Fig. \ref{PT_selective}(c) and (d)],
which show an anomalous non-monotonic behavior for the later case~[Fig. \ref{PT_selective}(d)]. That is, a state prepared in the center of a 2D lattice is first pumped to the top edge by bulk NHSE, then diffuses along the left edge due to the skin-free edge states.
Finally, the method has also been extended to 3D examples with other types of pseudo-Hermiticity (as a generation of $PT$ symmetry)~\cite{mostafazadeh2002pseudo}. In this paper~\cite{lei2023pt}, a rich variety of bulk, surface, and hinge NHSE has been engineered accordingly, where the later two can also be classified as HSTE or higher-order NHSE.

\subsection{HSTE with Hermitian bulk and non-Hermitian boundaries}
\begin{figure}[htbp]
\includegraphics[width=0.9\linewidth]{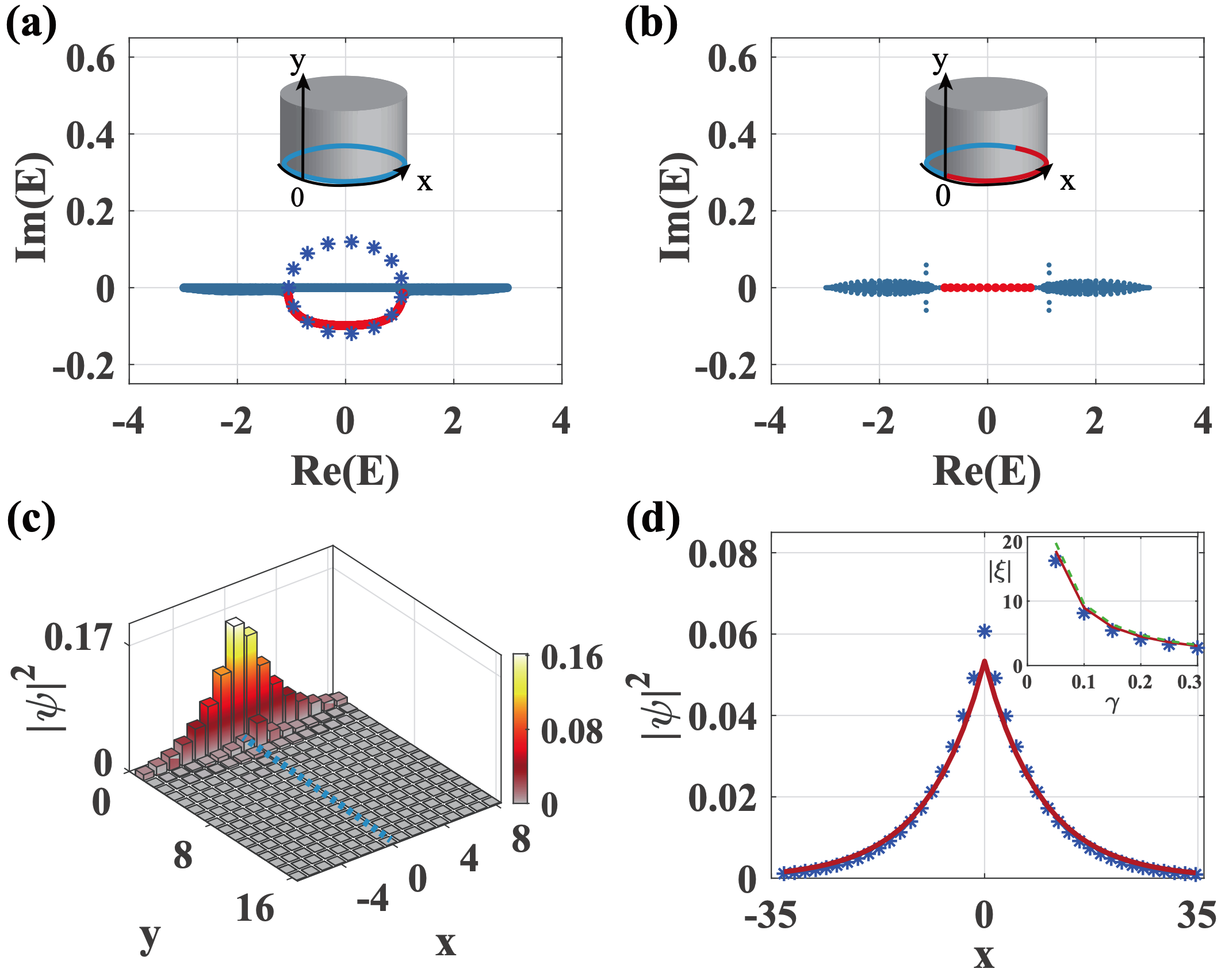}
\caption{Non-Hermitian chiral skin effect in a cylinder Haldane model with different edge dissipation.
(a) With a uniform dissipation acting on one of the two edges ($y=0$), the dissipative edge states form half of the loop-like spectrum (red) of a 1D Hatano-Nelson model (blue stars).
(b) Spectrum in the presence of a domain wall of the edge dissipation, where dissipative edge states acquire purely real eigenenergies.
(c) and (d) Particle distribution of topological edge states in the presence of the dissipation domain wall.
Reproduced from Ref. \cite{ma2023nonH}.
}
\label{NHCSE}
\end{figure}

A distinguished feature of HSTE is that bulk states are immune to NHSE and remain extended in the overall lattice.
Similar phenomena can also emerge in non-Hermitian systems with their bulk being Hermitian, or adiabatically connected to a Hermitian one, as recently been unveiled in several independent studies~\cite{ma2023nonH,schindler2023hermitian,daichi2023universal}.
In Ref. \cite{ma2023nonH}, the authors introduce a concept of non-Hermitian chiral skin effect, namely the NHSE of chiral edge modes in Chern insulators,  induced by domain walls of dissipation.
For demonstration, they consider the Haldane model in the presence of several different types of bulk or boundary dissipation.
A representative example is given in Fig. \ref{NHCSE}, where dissipation is added to a single outermost zigzag edge of the Haldane model in a cylinder geometry.
For uniform dissipation, It is found that the dissipative edge states can be mapped to half of the Hatano-Nelson model under PBCs, with their eigenenergies forming only the lower part of a complete loop, as shown in Fig. \ref{NHCSE}(a).
On the other hand, when a domain wall presences of the edge dissipation, chiral edge states possess real eigenenergies and become localized at the domain wall, as can be seen from Fig. \ref{NHCSE}(b) to (d).
The mechanism is further extended to the Haldane model with staggered on-site gain and loss, namely the model investigated in Ref.~\cite{li2022gain,zhu2022hybrid}, and provides an explanation for the emergence of HSTE therein.

Similar boundary NHSE has also been investigated in two contemporaneous works from different groups~\cite{schindler2023hermitian,daichi2023universal}, with their focuses diverging from each other.
In Ref. \cite{schindler2023hermitian}, the authors
propose a correspondence between Hermitian bulk topology and intrinsic non-Hermitian boundary topology in non-Hermitian systems with a line-gap. This correspondence is built on an important feature of topological phases, namely, topological properties shall remain unchanged during a continuous deformation that preserves the energy gap and certain symmetries of the system.
Therefore a line-gapped topological system with non-Hermitian perturbation can always deformed into a Hermitian limit, yet nontrivial non-Hermitian topology (spectral loop) and NHSE may still occur in its gapless topological boundary states.
Beyond the 2D Chern insulator characterized by an integer-valued topological invariant ($\mathbb{Z}$ invariant), Ref. \cite{schindler2023hermitian} has systematically investigate both $\mathbb{Z}$ and $\mathbb{Z}_2$ topological insulators in 2D and 3D, belonging to different classes of the Altland-Zirnbauer (AZ) symmetry classification for describing topological phases~\cite{AZ1997}.
On the other hand, Ref. \cite{daichi2023universal} emphasizes mainly on a universal platform for realizing non-Hermitian point-gap topology, by coupling a boundary of a topological insulator or superconductor to the environment and thus making it dissipative.
After introducing the general mechanism and demonstrating an example of the Chern insulator,
the authors extend their discussion to the full AZ$^\dagger$ symmetry classes, a modified version of the AZ classification based on symmetries intrinsic to non-Hermitian systems~\cite{kawabata2019symmetry}.

\section{Experimental realizations}\label{sec:exp}
\begin{figure*}
\includegraphics[width=0.75\linewidth]{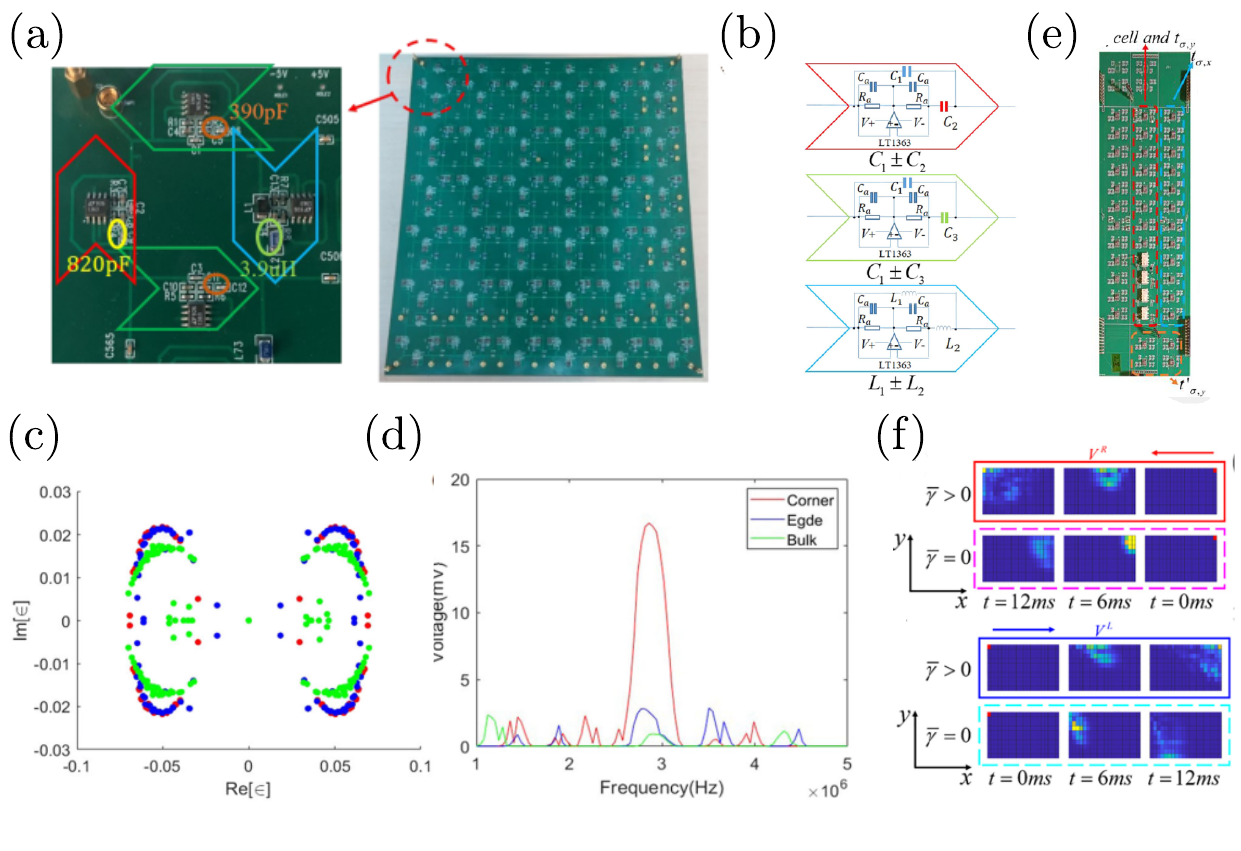}
\caption{
Experimental realization of hybrid skin-topological effect in electrical circuits system.
(a) The photograph of the fabricated electric circuits for the hybrid skin-topological effect induced by nonreciprocal couplings. Left presents the enlarged view of the unit cell. (b) Detail structure of different kinds of negative impedance converter through current inversion to realize the nonreciprocal couplings. (c) The measured spectrum for different boundary condition. Red, blue and green dots represent the spectra with full PBCs, x-PBCs/y-OBCs, and full OBCs respectively. (d) Measured voltage with the frequency for HSTE. (e) A sample of electric circuits to realize topological switch of non-Hermitian skin effect. (f) The electrical signal transport for different source excitation ($V_R$ or $V_L$) with the NHSE switched on or switched off. Reproduced from Refs.~\cite{zou2021observation,zhang2023electrical}.}
\label{experiment}
\end{figure*}

The experimental realization of HSTE is challenging for its high dimension, complexity of couplings and non-Hermitian instability. Up to now, it has been realized in a few metamaterial systems, including electric circuits system, non-Hermitian phototnic crystals, and active matter system, which will be reviewed in this section.
\subsection{HSTE in circuit systems}
The first realization of HSTE is given by Zou $et.$ $al.$ in an electric circuits system with non-reciprocal couplings~\cite{zou2021observation}. One sample is shown in Fig.~\ref{experiment}(a). Each unit cell contains four sites that are connected by capacitors, inductors and negative impedance converter through currents inversion as couplings. The negative impedance converter through currents inversion is used to realize non-reciprocal couplings [Fig.~\ref{experiment}(b)]. Thus two opposite non-reciprocal couplings are obtained in both $x$ direction and $y$ direction, and the net non-reciprocity of the bulk is zero. Things are different at boundary, where a set of non-reciprocal couplings is isolated due the topological localization, and the non-reciprocal pumping makes the waves localized at corners. Spectra for different boundary conditions can be obtained by the circuit Laplacian $J$, which is extracted by measuring the voltage $V_i^j$ with input currents $I_j$. The extracted spectra for different boundary conditions are shown in Fig.~\ref{experiment}(c). Parts of eigenvalues, whose number scales as $\mathcal{O}(L)$ with $L\times L$ the system's size, are quite sensitive to the boundary conditions.
In particular, a group of green points corresponding to eigenvalues of topological edge states under full OBCs are circled by blue points, which is one signature of HSTE. The voltages with frequency for HSTE are measured at corner, edge and bulk [Fig.~\ref{experiment}(d)], and most fields are seen to localize at corners.

A different realization of HSTE based on electrical circuits system is given by the same group in 2023~\cite{zhang2023electrical}, where they have mainly focused the application of HSTE as a topological switch of NHSE~\cite{li2020topological}.
The designed system is more complicated compared with the previous work, and a single unit cell is shown in Fig.~\ref{experiment}(e).
On the other hand, this setup does not require non-reciprocal coupling, and the non-Hermiticity is introduced solely by on-site loss.
The NHSE of the system can be switched on or off by changing the bulk topology of the system. As shown in Fig.~\ref{experiment}(f), in topological nontrivial case, the signal is non-reciprocally pumped to the corners along edges; in topological trivial case, the signal propagates into the bulk.

In another contemporaneous work, Shang $et.$ $al.$ have proposed an approach—physics-graph-informed machine learning (PGIML)—to process large amount of data in the experimental measurement~\cite{Shang2022},
and they also observe HSTE in an electrical circuits system through the PGIML.

\begin{figure*}[htbp]
\includegraphics[width=0.85\linewidth]{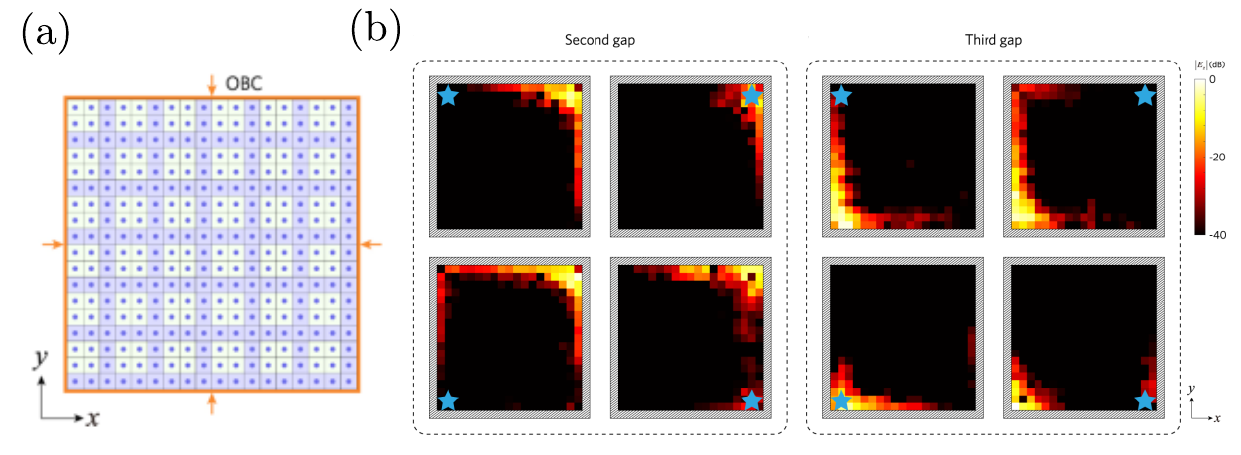}
\caption{
Experimental realization of hybrid skin-topological effect in non-Hermitian photonic crystals.
(a) Schematic diagram of the non-Hermitian gyromagnetic photonic crystal with designed loss. (b) Measured field distributions with different source excitation and different frequency (within second band gap or third band gap). Reproduced from \cite{Liu2023}.}
\label{phC}
\end{figure*}

\subsection{HSTE in non-Hermitian photonic crystal}
HSTE has also been observed In some very recent experimental developments of photonic systems~\cite{Liu2023,Sun2023floquet}.
In Ref. \cite{Liu2023},
Liu $et.$ $al.$ have realized HSTE in a non-Hermitian gyromagnetic photonic crystals~\cite{Liu2023}. The schematic diagram of the system is shown in Fig.~\ref{phC}(a), which is composed of gyromagnetic rods periodically arranged in a square lattice. Without loss, the system supports extended chiral edge states due to the nontrivial Chern topology. The loss is introduced by adding absorbing materials. With designed loss, the eigenvalues of topological edge states are changed from real to complex. For appropriately structured loss, the topological edge states along different edge acquire different imaginary parts for their eigenvalues, which then cover an area in the complex plane. Under x-PBCs/y-OBCs, different topological edge states are connected as a loop in their eigenvalues, dramatically different from those under double OBCs. The fields are seen to accumulate to corners that connect different edges, as shown in Fig.~\ref{phC}(b). For different excitation, the fields for the second band gap (third band gap) are localized to the right-up corner (left-down corner), reflecting the HSTE for edge states within the corresponding band gap.

HSTE has also been realized in non-Hermitian photonic Floquet lattices~\cite{Sun2023floquet}. Treating $z$ direction as time, the three dimensional photonic lattice can be used to simulate a two dimensional Floquet topological insulator with chiral edge states. Sun $et.$ $al.$ introduced structured loss into such a system to realize HSTE, where the one way edge states are imposed into specific corners. In addition, topological switch of NHSE was also observed by introducing a line-gap topological phase transition.

\subsection{HSTE in active matter system}
Active matter systems are a natural platform to study non-Hermitian topological physics, considering that these systems absorb and dissipate energy. In 2021, S. Palacios $et.$ $al.$ have proposed  a coupled-wire construction method to design microfabricated devices that support HSTE~\cite{Palacios2021}.
Their designed devices satisfy detailed balance in each unit cell, which is broken along top and bottom edges for topological nontrivial phases, manifesting as large corner accumulation for self-propelled Janus particles.


\section{Summary and outlook}
In this review, we have focused on the discovery and developments of HSTE,
starting from several typical examples with different non-Hemiticities and geometries,
to alternative theoretical comprehensions and engineering methods of the origin HSTE,
as well as its realizations in different experimental setups.
These investigations have already shown great potential in further exploration of HSTE, not only in the phenomenon itself, but also its interplay with other fascinating physics in different research fields, and here we shall list some possible directions.
(i) The various mechanisms of HSTE discussed in this review already call for a systematic classification of these phases, and distinct properties of different classes of HSTE are to be explored.
(ii) As a boundary phenomena, variations of HSTE may also arise in higher-dimensional non-Hermitian systems with different defects or impurities, while investigations along this route has only focused on bulk properties so far~\cite{disclination_NHSE,dislocation_NHSE1,discloation_NHSE2,dislocation_NHSE3,manna2023inner}.
(iii) The destructive interference of non-reciprocity (which generates HSTE in Ref. \cite{Lee2019hybrid}) has also been shown to induce some anomalous phenomena in 1D, e.g., the critical NHSE~\cite{li2020critical,liu2020helical,yokomizo2021scaling,qin2023universal} and direction reversal of NHSE~\cite{li2022direction}, which have not yet be explored in the context of HSTE in higher dimensions.
(iv) Many-body physics also plays an important role in various nontrivial phases in condensed matter physics.
Its applications for non-Hermitian systems are mostly on 1D systems, which already lead to exotic emergent physics in multi-particle systems~\cite{mu2020emergent,lee2021many,shen2022non,faugno2022interaction,orito2022unusual,qin2023occupation}.
For higher-dimensional systems,
the skin accumulation of HSTE can be independently tuned for bulk and boundaries of different dimensions,
thereby providing a platform for investigating the interplay between skin accumulation, many-body physics, and dimensionality within a unified system.
We note that these are only a few directions that in our view are most closely related to the idea of HSTE,
and many other possibilities are yet to be discovered.

\section{Acknowledgement}
Linhu Li acknowledges support from the National Natural Science Foundation of China (Grant No. 12104519),
the Guangdong Basic and Applied Basic Research Foundation (Grant No. 2020A1515110773),
and the Guangdong Project (Grant No. 2021QN02X073).

\bibliography{nonH_hybrid_rev}
\end{document}